\documentclass[letterpaper,spanish]{IEEEtran}
\usepackage[T1]{fontenc}
\usepackage[latin9]{inputenc}
\usepackage{color}
\usepackage{babel}
\addto\shorthandsspanish{\spanishdeactivate{~<>}}

\usepackage{array}
\usepackage{varioref}
\usepackage{textcomp}
\usepackage{graphicx}
\usepackage[unicode=true,
 bookmarks=true,bookmarksnumbered=false,bookmarksopen=false,
 breaklinks=true,pdfborder={0 0 1},backref=false,colorlinks=true]
 {hyperref}
\hypersetup{pdftitle={El Wiimote en el salón de clases},
 pdfauthor={Eduardo NAVAS},
 pdfsubject={Uso del wiimote en el salón de clases},
 pdfkeywords={wii wiimote nintendo clases tecnología táctil}}

\makeatletter

\pdfpageheight\paperheight
\pdfpagewidth\paperwidth

\providecommand{\tabularnewline}{\\}

\makeatother

\usepackage{listings}
\begin{document}
\title{Tecnología Portátil de Pantalla Táctil de Bajo Costo, Aplicada a la
Docencia Universitaria\thanks{Este artículo fue presentado en el \textbf{Congreso de Electrónca
e Informática 2010} de la \emph{Universidad Centroamericana ``José
Simeón Cañas''}, celebrado el 25 y 26 de noviembre de 2010.}}
\author{Eduardo Adam NAVAS-LÓPEZ, Catedrático del Departamento de Electrónica
e Informática, Universidad Centroamericana ``José Simeón Cañas'',
enavas@ing.uca.edu.sv}

\maketitle

\begin{abstract}
En este artículo se describe una implementación de tecnología portatil
de pantalla táctil de bajo costo, aplicada a la docencia universitaria,
utilizando como base el control remoto de la consola Wii de Nintendo
(conocido como Wiimote), un proyector de cañón normal, una computadora
y software libre. El propósito es mostrar la viabilidad de dicha implementación
para mejorar los procesos de enseñanza/aprendizaje, sin incurrir en
altos costos asociados a equipo tecnológico no asequible, a infraestrutura
especial en los salones de clase, o a programas de computadora costosos.
Se incluye también un resumen de una prueba del sistema en dos cursos
universitarios.
\end{abstract}


\section{Introducción}

\IEEEPARstart{C}{omo} docente en un país empobrecido y con altos
niveles de desigualdad como El Salvador, se puede constatar que la
calidad de los procesos de enseñanza-aprendizaje está fuertemente
supeditada a la disponibilidad o escasez de recursos de los centros
de estudio y a las condiciones socioeconómicas de los profesores y
los alumnos.

Y como docente y como informático, también se puede constatar que
las Tecnologías de Información y Comunicación (TICs) disponibles en
la actualidad, permiten dinamizar y potenciar muchas estrategias novedosas
en el campo de la docencia. Sin embargo, muchas de estas tecnologías
son desconocidas por los docentes, no están al alcance económico de
estos incluso en su forma más básica, o si lo están, se desaprovecha
su potencial por falta de capacitación apropiada. Conversando con
alumnos de grado en la universidad que cursan materias relacionadas
con Programación de computadoras y TICs, se constata que la mayoría
de ellos tuvo profesores (en su educación básica y media) que no tenían
ningún tipo de destreza en TICs.

Aquellos que sí tenemos capacitación/conocimientos/formación en TICs,
sabemos que existen diversas herramientas para diferentes propósitos.
Algunas son muy costosas y otras muy económicas, algunas son tecnologías
propietarias y cerradas\footnote{Por tecnología \emph{propietaria}, se entiende una tecnología que
es propiedad de una empresa privada, y por tecnología \emph{cerrada},
se entiende una tecnología (generalmente propietaria) cuya especificación
se mantiene en secreto dentro de una empresa que la controla y dirige
su desarrollo. Ejemplos de tecnologías propietarias y cerradas son
los sistemas operativos Windows, Mac Os X, los teléfonos iPhone, etc.}, y otras son tecnologías libres y abiertas\footnote{Las tecnologías libres y abiertas generalmente son de acceso gratuito,
pero su principal característica es que su especificación está disponible
para cualquiera que desee modificarlas y adaptarlas a sus necesidades
particulares. Ejemplos de tecnologías libres y abiertas son los sistemas
operativos GNU/Linux, los programas del proyecto GNU, el formato multimedia
OGG, algunos lenguajes de programación como Python y Lua, etc.}. Es por ello que este artículo pretende contribuír a acercar tecnologías
libres, abiertas y de bajo costo económico a la población docente,
principalmente universitaria, aunque aplicable a otros niveles educativos.

\section{Metodología}

\subsection{\label{subsec:Alcances}Alcances de la implementación}

\IEEEPARstart{E}{ste} artículo explica cómo implementar una pantalla
táctil del tamaño de una pizarra mediana (de hasta $150\times110cm^{2}$)
para controlar una computadora.

Lo que se logra con esta implementación es que, al proyectar el escritorio
de la computadora sobre una superficie plana al frente del salón de
clase (o a un costado), se puede ``controlar'' la computadora (es
decir, controlar el Sistema Operativo) desde el lugar donde se proyecta,
sin tener que manipular físicamente la computadora, sin necesitar
un asistente que controle la computadora mientras el profesor explica
lo que está proyectado, y sin que el profesor tenga que desplazarse
continuamente entre la computadora y la superficie de proyección.

\begin{figure}
\begin{centering}
\includegraphics[width=1\columnwidth]{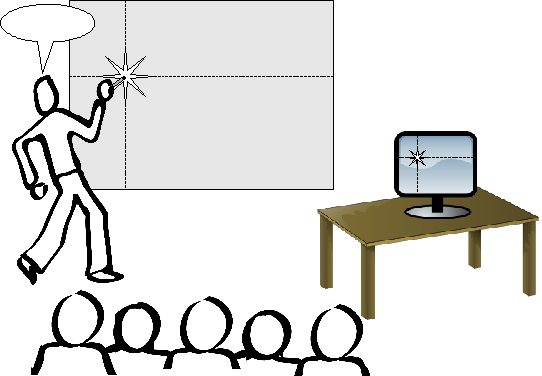}
\par\end{centering}
\caption{\label{fig:profesor+sistema}El clic sobre la superficie de proyección
se convierte en clic en la computadora}

\end{figure}

Esta implementación provee una buena ``emulación'' del ratón de
la computadora, desde la superficie de proyección, con ayuda de un
pulsador infrarrojo (cuya construcción se describe en la sección \ref{subsec:Construcci=0000F3n-pulsador}).
Es decir, que al activar el pulsador infrarrojo sobre la superficie
de proyección, se genera el evento de un clic (izquierdo, derecho,
central, etc.) en la correspondiente ubicación en la pantalla de la
computadora (ver figura \ref{fig:profesor+sistema}).

Los sistemas operativos modernos permiten un uso y control\footnote{Entendiéndose el tipo de uso y control de un usuario normal de escritorio.}
casi total de sí mismos, utilizando únicamente un ratón con tres botones
y un dispositivo de rodamiento unidimensional bidireccional (o \emph{scroll
wheel}). Por ello, realizar operaciones comunes, como abrir archivos,
guardar archivos, moverse entre las carpetas del sistema de archivos,
cambiar el foco de atención de una ventana a otra, desplegar menús,
seleccionar herramientas de las barras, copiar archivos, etc., son
operaciones que se pueden realizar fácilmente y con cierta comodidad
únicamente usando un ratón estándar moderno.

Por otro lado, si se utilizan aplicaciones (programas) especiales
para depender poco, o nada, de un teclado completo para impartir las
clases, esta implementación se potencia enormemente.

Hay diversas formas de impartir clases, y las computadoras se introducen
en ellas de diferente manera, en el salón de clase. A continuación
se mencionan las identificadas en la experiencia personal del autor
y que son relevantes en este contexto: Una de ellas, es impartir las
clases a partir de presentaciones de diapositivas electrónicas. Si
el lector ha utilizado diapositivas electrónicas en sus clases (como
profesor o como alumno), habrá notado lo estática que es la interacción.
También habrá notado que se depende totalmente de que las diapositivas
estén bien preparadas. Por ejemplo, si un alumno hace una pregunta
que no puede contestarse con la presentación, el profesor debe recurrir
a una pizarra para responder a la pregunta. Con la implementación
presentada aquí, la pantalla de la computadora es la pizarra, y el
profesor simplemente tendría que cambiarse a otra aplicación, para
escribir la explicación de la pregunta.

Otra forma de impartir clase, en la que la computadora se suele usar,
es para hacer demostraciones de programas de computadora que ilustran
los conceptos que se quieren transmitir, o simplemente ejemplificar
lo que los alumnos deben hacer para usar ellos mismos los programas
en cuestión. Estas demostraciones normalmente exigen que el profesor
o instructor, esté ``pegado'' a la computadora durante la clase;
y cuando necesita señalar algo en la pantalla, debe recurrir a una
de tres cosas: señalar con el puntero del ratón (lo cual es bastante
impersonal), señalar con un puntero laser (lo cual implica equipo
adicional) o desplazarse físicamente hacia la superficie de proyección
y señarlar con el dedo o la mano. Una forma de agilizar este ir y
venir entre la computadora y la superficie de proyección es con un
asistente que controle la computadora, pero eso aumenta el personal
requerido.

Por otro lado, en la forma tradicional de impartir clases (especialmente
en temáticas de corte humanístico), lo que se necesita es simplemente
escribir texto, líneas y algunos diagramas simples en la pizarra,
al momento de la clase. Si surge una pregunta, el profesor no está
limitado de escribir más en la pizarra para responder. Aunque sí está
limitado por el espacio de la pizarra, y tal vez tenga que borrar
para poder escribir más. Para ello, la computadora y el proyector
más bien estorban, ya que normalmente, no se puede guardar en ella,
lo que se escribe en la pizarra.

Con la implementación presentada aquí, estos tres tipos de técnicas
didácticas pueden mesclarze en una misma clase o cátedra. Se pueden
utilizar presentaciones en diapositivas, pero el profesor no necesita
tomar un plumón o yeso para responder preguntas inesperadas, porque
puede escribir sobre la superficie de proyección. Se pueden hacer
demostraciones de programas directamente desde la superficie de proyección
sin tener que desplazarse continuamente hacia y desde la computadora.
Y por supuesto, lo más importante, es que se pueden realizar las clases
tradicionales, en las que se explican conceptos escribiendo ``en
la pizarra''. Además de eso, estos garabatos, textos, líneas, diagramas,
etc., se pueden guardar en la computadora, y servir después como referencia
para los alumnos y para el profesor.

Por supuesto hay que considerar también todas las limitaciones ambientales
de un proyector de cañón, como su luminosidad y su contraste en un
determinado salón de clase, con una determinada iluminación. También
hay que considerar el tamaño de la clase, es decir, el número de alumnos,
porque una pantalla de $140\times100cm^{2}$ puede ser demasiado pequeña
para 100 o más alumnos, dependiendo de la forma del aula y la distribución
de los alumnos.

\subsection{Requerimientos de esta implementación}

\begin{figure}
\begin{centering}
\includegraphics[width=1\columnwidth]{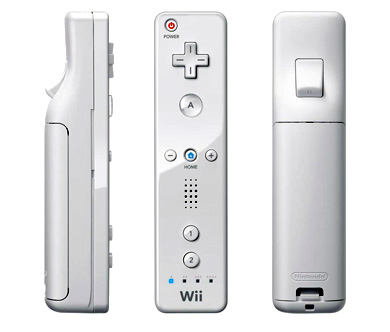}
\par\end{centering}
\caption{\label{fig:wiimote}Vistas del Control remoto de la consola Wii de
Nintendo, conocido como \emph{wiimote}}
\end{figure}

En general, los componentes físicos (hardware) necesarios son:
\begin{itemize}
\item Una computadora (laptop, desktop o netbook\footnote{También conocidas como ``Mini Laptop'', tienen como característica
principal, sus pequeñas dimensiones y su falta de unidad óptica.}), con transmisor bluetooth (de clase 1 o clase 2) que puede ser interno
o externo,
\item un proyector de cañón (con resolución mínima de $1024\times768$\footnote{En realidad funciona con cualquier resolución, pero esta es una bastante
cómoda, ya que no es ni muy pequeña, ni muy grande.}),
\item un control remoto de la consola Wii de Nintendo, conocido como \emph{Wiimote}
(ver figura \ref{fig:wiimote}), y 
\item un pulsador infrarrojo (de fácil fabricación).
\end{itemize}
El componente \emph{wiimote} cuenta con una serie de dispositivos
de entrada y salida de información:
\begin{enumerate}
\item Una serie de botones que siguen la tradición de los controles de Nintendo.
\item Cuatro LEDs para indicar al usuario qué jugador es (1-4) durante los
juegos (la consola \emph{Wii} soporta entre 1 y cuatro jugadores simultaneamente).
\item Una memoria para almacenar información de los juegos.
\item Un vibrador para usarse con los juegos.
\item Tres acelerómetros (es decir, medidores de aceleración). Uno por cada
eje del control remoto.
\item Un transmisor Bluetooth.
\item Una cámara infrarroja en el frente.
\end{enumerate}

A continuación se detallan más los requerimientos mínimos a considerar
para la implementación:

\subsubsection{En el aula}

En el salón de clase únicamente se necesita contar con un tomacorriente
estándar, de preferencia polarizado para proteger al proyector, y
una superficie limpia, lisa y de color claro\footnote{Lo ideal es que sea blanca, pero los colores ``blanco-huezo'' o
similares están bien.} y uniforme de unos $2\times1,5m^{2}$.

\subsubsection{Componentes físicos electrónicos (hardware)}

\begin{figure}
\begin{centering}
\includegraphics[width=0.75\columnwidth]{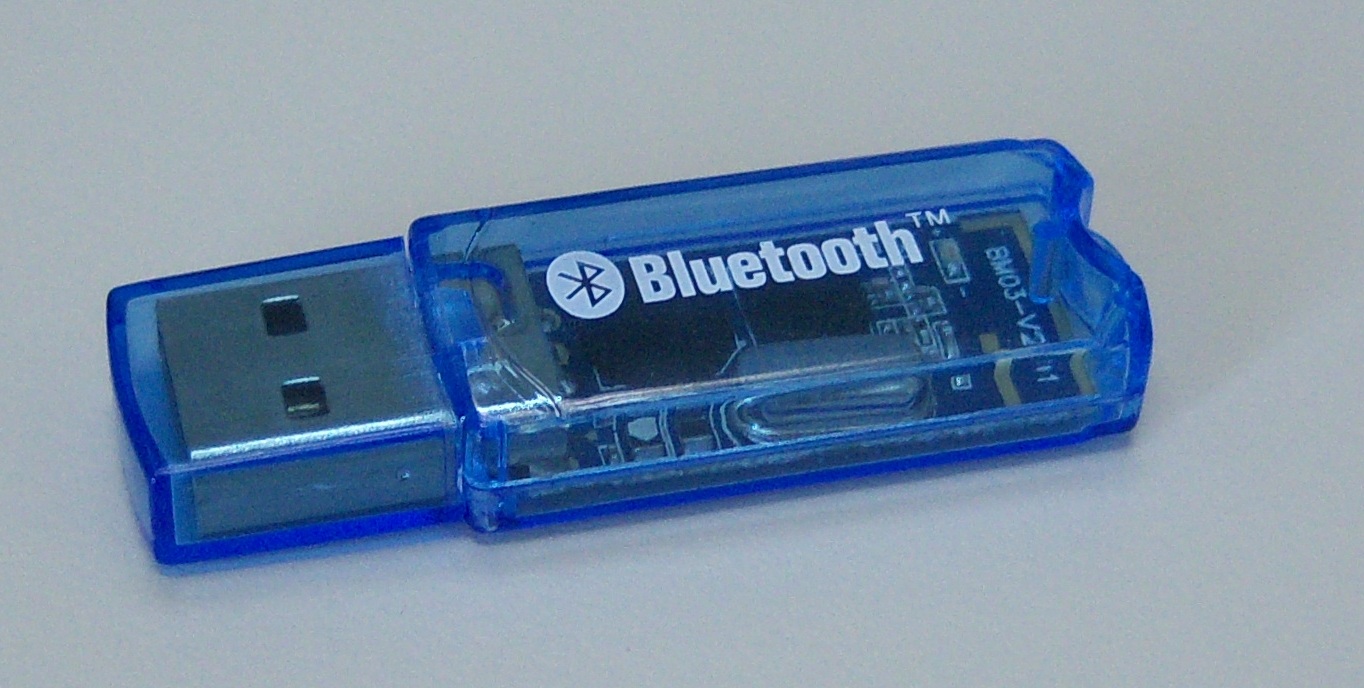}
\par\end{centering}
\caption{\label{fig:Bluetooth}Un transmisor Bluetooth USB}

\end{figure}

Tal como se mencionó arriba, se necesita un proyector de cañón con
resolución recomendada de $1024\times768$ (que es la típica para
proyecciones), un control remoto\footnote{Se recomienda fuertemente que se use con la funda antigolpes propia
del control, para salvaguardarlo de caídas. Esta funda viene con el
control cuando se compra por separado. Puede verse en la figura \vref{fig:Wiimote-montado}.} de la consola Nintendo Wii (no se necesita la consola, sólo el control
remoto, que es un accesorio que se vende por separado), un pulsador
infrarrojo de mano y una computadora.\\
La construcción del pulsador infrarrojo se describe en la subsección
\ref{subsec:Construcci=0000F3n-pulsador}.\\
Se necesita que la computadora tenga transmisor Bluetooh, ya sea interno
o externo (como el de la figura \ref{fig:Bluetooth}).\\
La computadora puede ser un equipo con bajos recursos. Hay distribuciones
GNU/Linux especializadas en equipos así, como \href{http://www.xubuntu.com/}{Xubuntu},
\href{http://spins.fedoraproject.org/lxde/}{Fedora spin LXDE}, \href{http://spins.fedoraproject.org/xfce/}{Fedora spin XFCE},
etc. Además, hay muchas distribuciones GNU/Linux que se pueden configurar
manualmente para requerir muy pocos recursos, como \href{http://www.debian.org}{Debian},
\href{http://www.archlinux.org/}{Arch Linux}, \href{http://www.slackware.org/}{Slackware},
etc. Léase, por ejemplo, las características de las computadoras utilizadas
en las pruebas, que se describen en la sección \ref{subsec:Prueba-real}.

\subsubsection{Componentes lógicos (software)}

Los componentes lógicos o programas de computadora (software) necesarios,
que además son programas libres, son:
\begin{itemize}
\item Un sistema operativo GNU/Linux, de preferencia una distribución \emph{{*}ubuntu
9.04} o superior,
\item el programa \texttt{wmgui} para determinar la alineación correcta
del wiimote, y
\item el programa \texttt{gtkwhiteboard}, para calibrar el wiimote con la
superficie de proyección.
\item Además, cualquier otro programa necesario y/o útil para las clases
específicas (véase el apéndice \vref{sec:otros-programas}).
\end{itemize}

\subsection{Funcionamiento}

El funcionamiento del sistema se explica a continuación, y como se
verá, el centro de todo es el wiimote (ver figura \vref{fig:wiimote}).

\begin{figure}
\begin{centering}
\includegraphics[width=0.5\columnwidth]{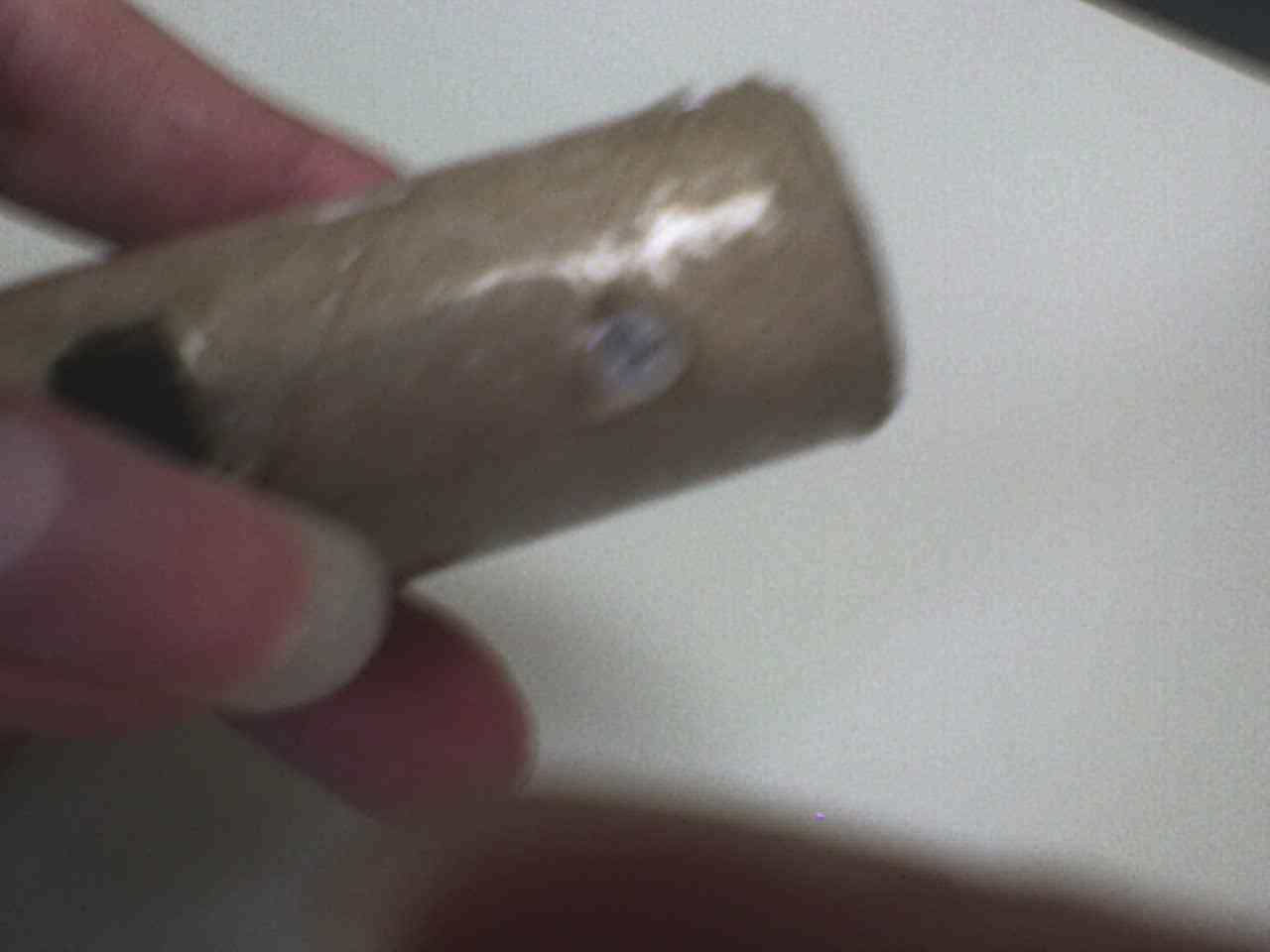}\includegraphics[width=0.5\columnwidth]{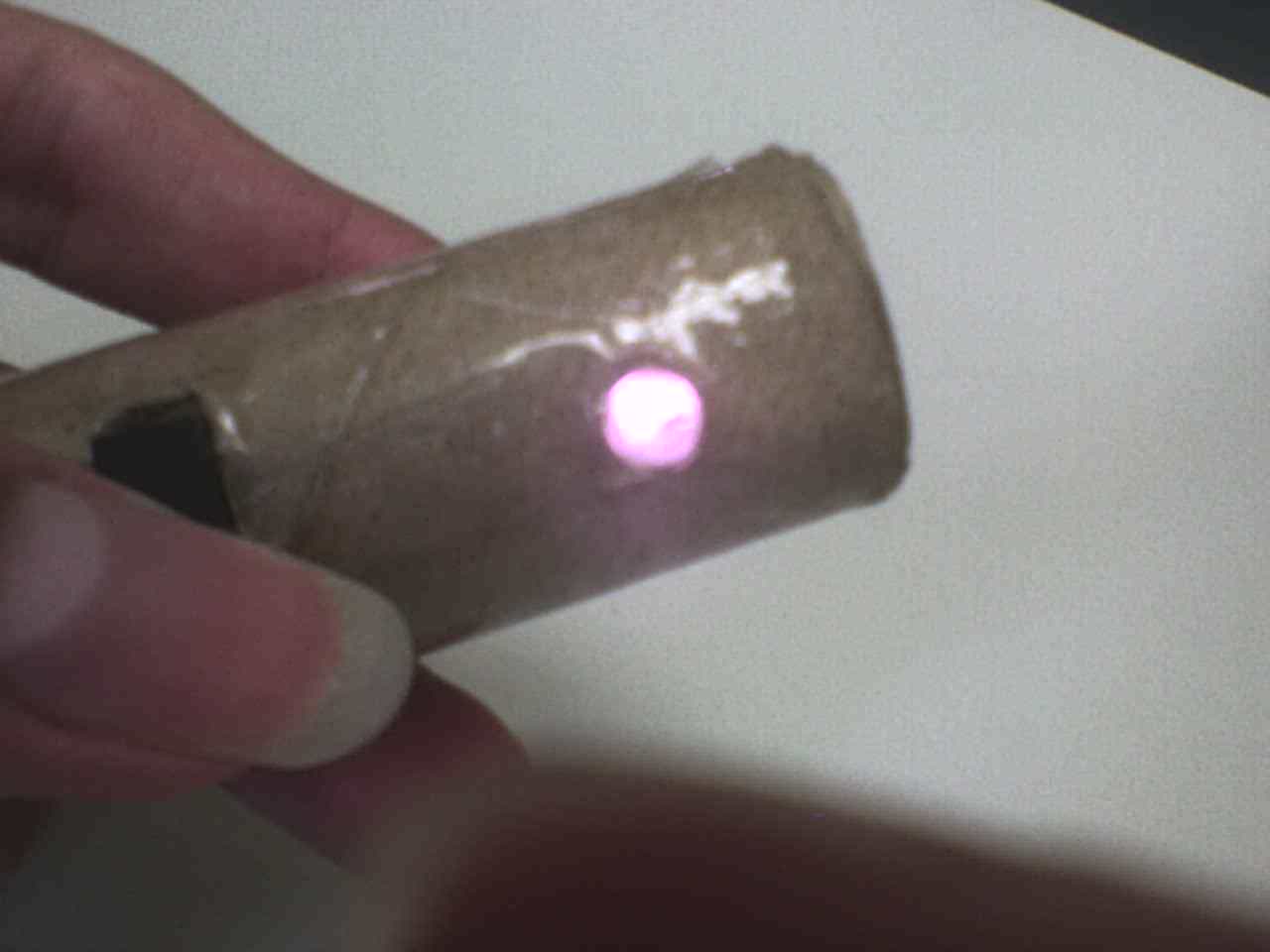}
\par\end{centering}
\caption{\label{fig:Pulsador-infrarrojo}Pulsador infrarrojo apagado y encendido,
visto a através de una cámara de celular.}

\end{figure}

Como se dijo anteriormente, el wiimote es un control remoto para la
consola de videojuegos Wii de Nintendo, y para comunicarse dispone
de un transmisor bluetooth y una cámara de detección infrarroja multipunto.
La cámara infrarroja no capta imágenes infrarrojas térmicas como las
que usa la policía de algunos países, sino que capta puntos de radiación
electromagnética infrarroja. Esta radiación infrarroja (o luz infrarroja),
que es invisible para los humanos, puede verse a través de la mayoría
de cámaras digitales y cámaras de celular, como lo ilustra la figura
\vref{fig:Pulsador-infrarrojo}.

\begin{figure}
\begin{centering}
\includegraphics[width=1\columnwidth]{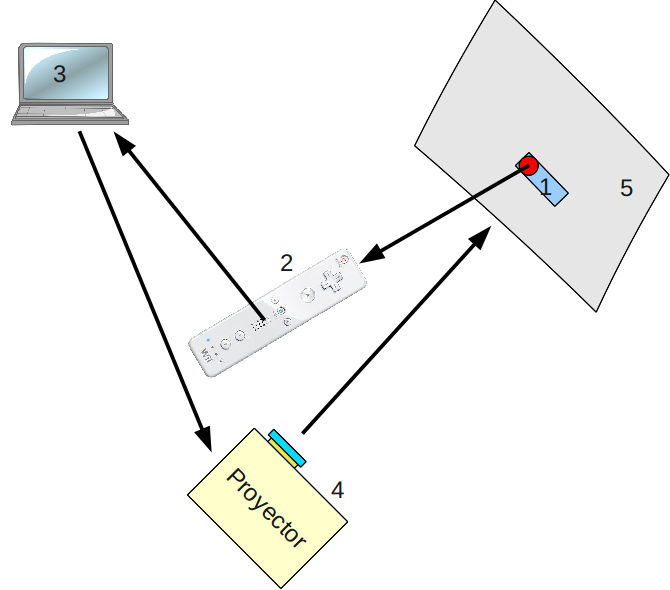}
\par\end{centering}
\caption{\label{fig:Esquema-de-funcionamiento}Esquema de funcionamiento\protect \\
1.- Pulsador infrarrojo\protect \\
2.- Wiimote\protect \\
3.- Programa de comunicación en la computadora\protect \\
4.- Proyector de cañón\protect \\
5.- Superficie de proyección}

\end{figure}

Entonces, una vez configurado el sistema, el usuario (el profesor)
activa el pulsador infrarrojo cerca de la superficie de proyección
cuando quiere hacer clic; la cámara infrarroja del wiimote capta el
destello y retransmite la imagen a la computadora por el transmisor
bluetooth; luego la computadora, a través de un programa especial
para tal efecto, transforma la información transmitida de la cámara
en una coordenada de la pantalla, donde generar el evento de clic.
Esto se ilustra en la figura \vref{fig:Esquema-de-funcionamiento}.

\subsection{Preparación previa al uso en clases}

Debe aprenderse a usar todo el sistema y configurarlo previamente
antes de poder usarlo. A continuación se presentan los pasos que deben
seguirse:
\begin{enumerate}
\item Asegurarse de contar con los componentes físicos necesarios, y que
estos funcionen adecuadamente. El pulsador infrarrojo puede probarse
viendo el led infrarrojo a través de una cámara digital de celular
o de video. El wiimote puede probarse presionando el botón rojo que
está en el compartimiento de las baterías, al hacerlo, las luces frontales
comienzan a parpadear.\\
Las aplicaciones de conexión que se describen en la sección \ref{subsec:Puesta-en-marcha},
indican el nivel de la batería (en porcentaje). En el experimento
descrito en la sección \ref{subsec:Prueba-real} se identificó que
el porcentaje mínimo funcional de las baterías en el wiimote para
poder usar la cámara infrarroja es del 50\%.
\item Instalar los programas necesarios.\\
Si el sistema operativo elegido es un {*}ubuntu\,9.04 o superior,
la instalación de los programas de comunicación con el wiimote se
realiza con los siguientes comandos:\\
\texttt{\$ sudo apt-get install wmgui}~\\
\texttt{\$ sudo apt-get install gtkwhiteboard}\\
Se pueden también instalar algunos programas propios para su utilización
en clases como los descritos en el apéndice de la sección \ref{sec:otros-programas},
para lo cual se puede ejecutar los siguientes comandos:\\
\texttt{\$ sudo apt-get install xournal}~\\
\texttt{\$ sudo apt-get install cellwriter}~\\
\texttt{\$ sudo apt-get install dasher}~\\
\texttt{\$ sudo apt-get install kvkbd}
\item Se recomienda constuír un soporte para colocar el wiimote sobre el
proyector de cañón, a menos que se cuente con un pedestal y soporte
para micrófono o videocámara, o algo similar. En el apéndice \vref{sec:Ap=0000E9ndice-plataforma}
se sugiere un diseño de soporte.
\item Hacer una prueba inicial de conexión, en las mismas condiciones físicas
que las reales de utilización (además de probar las aplicaciones que
se utilizarán durante las clases, por supuesto). Para ello deben seguirse
los pasos de la sección \ref{subsec:Puesta-en-marcha}.
\item Al finalizar, o durante la prueba inicial de conexión, configurar
las aplicaciones (los programas) a utilizar durante las clases, para
que su uso sea fluido durante las mismas.\\
Un punto importante a determinar es la resolución apropiada para la
proyección. Debe, lógicamente, ser una resolución con la que tanto
el profesor como los alumnos se sientan cómodos y que además permita
transcurrir la clase con normalidad y comodidad.
\end{enumerate}

\subsection{\label{subsec:Puesta-en-marcha}Puesta en marcha al inicio de cada
clase}

Cada vez que se va a iniciar la clase, se deben seguir ciertos pasos
para la inicialización del sistema (la configuración de las aplicaciones
a usar debe haberse hecho previamente). Dichos pasos son los siguientes:
\begin{enumerate}
\item Primero debe iniciar la computadora y configurar la transmisión de
imagen al proyector de cañón.\\
Nota importante: El sistema no funciona apropiadamente con dos o más
pantallas activas al mismo tiempo, por lo que se debe elegir entre
espejar la misma imagen en la pantalla de la computadora y el proyector
de cañón, o desactivar la pantalla de la computadora y dejar sólo
la imagen del proyector de cañón. Esta última alternativa es la necesaria
en una netbook (típicamente con resolución $1024\times576$) y el
arreglo puede verse en la figura \vref{fig:Arreglo-de-proyector-en-netbook}.
\item Luego, debe colocarse el wiimote a una distancia cercana a, y apuntando
hacia, la superficie de proyección. Usualmente esta ubicación es a
un lado del proyector o encima de él (ver la figura \vref{fig:Wiimote-montado}).
Sin embargo, hay que tener en cuenta que los proyectores de cañón
generan mucho calor y hay que mantener al wiimote aislado térmicamente
de él, para evitar que también se sobrecaliente.
\item Ejecutar el programa \texttt{wmgui} para alinear correctamente el
wiimote (ver figura \ref{fig:wmgui}).
\begin{figure}
\begin{centering}
\includegraphics[width=1\columnwidth]{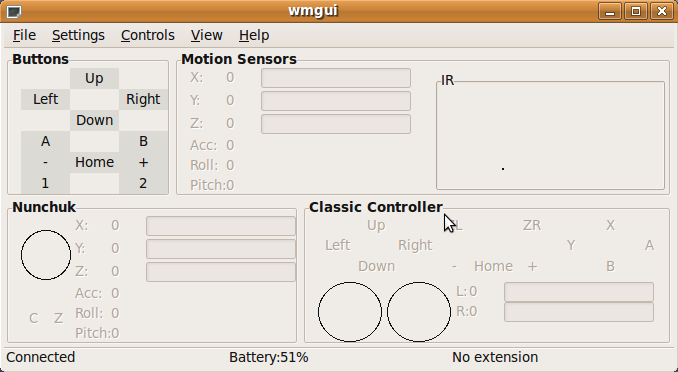}
\par\end{centering}
\caption{\label{fig:wmgui}Aplicación \texttt{wmgui} para alineación del \emph{wiimote}}

\end{figure}

\begin{enumerate}
\item Primero hay que conectar la aplicación con el wiimote, con la combinación
\textsf{CTRL+C}.
\item Presionar los botones \textsf{A} y \textsf{B} del wiimote simultaneamente.
\item Aceptar el aviso que muestra la aplicación (con \textsf{Intro}).
\item Una vez conectado, hay que activar la lectura de la cámara infrarroja,
con \textsf{CTRL+I}.
\item Luego, viene la parte más importante: Se debe garantizar que el pulsador
infrarrojo sea visible por la cámara del wiimote (lo cual puede verse
en el marco IR de la aplicación) en las cuatro esquinas de la imagen
proyectada. De no ser así, debe reubicarse el wiimote, o el proyector
para lograrlo (o bien, ajustar la imagen con las propiedades del proyector).
En la figura \ref{fig:wmgui} se ve un pequeño punto negro en el marco
IR que indica que el pulsador infrarrojo es visible.\\
\item Una vez garantizado que las cuatro esquinas de la imagen proyectada
son visibles por la cámara, la aplicación debe desconectarse del wiimote
(con \textsf{CTRL+D}) y luego cerrarse (con \textsf{ALT+F4}).
\end{enumerate}
\item Ejecutar la aplicación de sincronización \texttt{gtkwhiteboard} (o
su versión en línea de comandos, \texttt{cliwhiteboard}) para convertir
los destellos infrarrojos del pulsador, en clics para el sistema operativo.
\begin{figure}
\begin{centering}
\includegraphics[width=0.8\columnwidth]{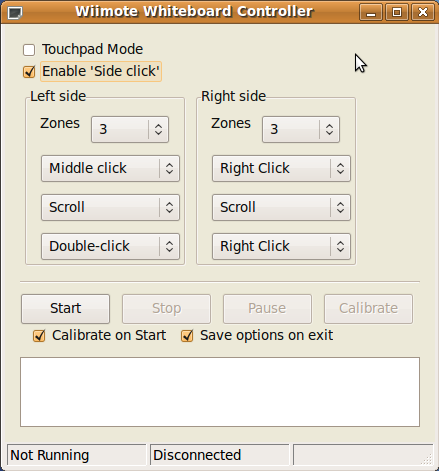}
\par\end{centering}
\caption{\label{fig:gtkwhiteboard}Aplicación de sincronización \texttt{gtkwhiteboard}
para la conexión con el \emph{wiimote}}

\end{figure}

\begin{enumerate}
\item Antes que nada, desactivar el modo Touchpad.
\item Luego se debe habilitar el ``clic lateral'' si se desea poder emitir
clic derecho, central, doble clic, etc.
\item En caso de necesitar clic derecho durante la clase, deben configurarse
las ``zonas'' de los bordes laterales (y probablemente sea buena
idea configurarlas siempre).
\begin{figure}
\begin{centering}
\includegraphics[width=0.75\columnwidth]{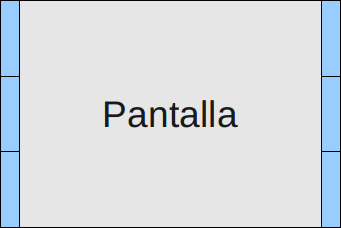}
\par\end{centering}
\caption{\label{fig:Zonas-laterales}``Zonas laterales'' para la pizarra}

\end{figure}
\\
A cada lado de la superficie de proyección, se pueden configurar hasta
tres zonas diferentes, tales que, al encender el pulsador infrarrojo
en ellas, equivale a realizar la acción indicada.\\
\\
Por ejemplo, si se configuran estas zonas laterales tal como en la
figura \vref{fig:gtkwhiteboard}, significa que la superficie de proyección
se segmenta como en la figura \vref{fig:Zonas-laterales}.\\
Cuando se enciende el pulsador infrarrojo dentro de la superficie
``Pantalla'', se genera un clic en la coordenada correspondiente
en la computadora. Si se enciende el pulsador infrarrojo uno o dos
centímetros a la derecha de la superficie de proyección en el tercio
superior de la altura, se genera un evento de clic derecho en donde
está el cursor del ratón en ese momento.\\
Cuando se enciende el pulsador uno o dos centímetros a la izquierda
de la superficie de proyección en el tercio inferior de la altura,
se genera un evento de doble clic en donde está el cursor del ratón
en ese momento; y así sucesivamente con las demás zonas.
\item Para no tener que indicar estas configuraciones cada vez, hay que
indicarle a la aplicación que guarde las opciones con la opción ``Save
options on exit''.
\item A continuación, hay que calibrar el wiimote con la pantalla, presionando
el botón ``Start''.\\
Hay que encender el pulsador en las cuatro esquinas de la pantalla,
siguiendo un patrón de ``\texttt{Z}''.
\end{enumerate}
\item Cuando la aplicación reporta que la calibración está completa (con
el mensaje ``Calibration complete!''), el pulsador ya puede usarse
para generar eventos de clic en la computadora.
\item Una vez que la clase haya concluído, hay que presionar el botón ``Stop''
o simplemente cerrar la ventana de la aplicación.
\end{enumerate}

\subsection{\label{subsec:Prueba-real}Prueba real en clases de programación
de computadoras}

Para verificar que efectivamente, la implementación se puede usar
eficazmente en clases, y en particular en clases universitarias, el
autor probó el sistema en dos cursos universitarios en el área de
programación de computadaras que imparte en el primer semestre de
2010.

A continuación se describen las condiciones en las que el sistema
se usó:

El primero de los cursos es sobre \emph{Programación Funcional}, en
el que se enseña a los alumnos un nuevo paradigma de programación
y un nuevo lenguaje de programación que tiene un IDE\footnote{\emph{Integrated Development Environment}, o Entorno de Desarrollo
Integrado, es un programa de computadora que sirve para hacer programas
de computadora.} interactivo, que es con el que se programa. Mucho del material base
está en formato PDF, aunque en forma de prosa estructurada, sin ilustraciones.
La dinámica seguida es la explicación de los conceptos nuevos, seguido
de una ejemplifiación de programación en el IDE correspondiente.

El segundo es sobre \emph{Graficación por Computadora}, en el que
se usa un libro de texto que los alumnos tienen completamente a disposición
en formato PDF, ya que es un libro libre (\href{http://dei.uca.edu.sv/publicaciones/libro_graficos_v1.0.tar.gz}{http://dei.uca.edu.sv/publicaciones/libro\_graficos\_v1.0.tar.gz}).
Este libro contiene abundantes ilustraciones. En este curso, se explican
muchos conceptos sobre geometría y matemática aplicada a la programación
de computadoras. Además, se presentan demostraciones de programas
de computadora que aplican y ejemplifican los conceptos.

En ambos casos, aunque casi todos los conceptos explicados en la clase
están en prosa en los archivos PDF, algunas explicaciones, como en
cualquier área de conocimiento, resultan más claras con diagramas
que con texto en prosa. Y además hay ocaciones en las que, aunque
los diagramas sí estén presentes en el material escrito, es pedagógicamente
más apropiado explicarlos construyéndolos, que explicarlos ya terminados.

Para las clases se contó con salones elípticos, en los que cada alumno
está en un pupitre. Hay una pizarra de yeso al frente y hay tres bloques
de lámparas fluorescentes que se pueden apagar/encender independientemente,
así como varios tomacorrientes trifilares sin respaldo de energía.
Cada salón, también tiene una pantalla de proyección retráctil que
se puede extender justo en el medio de la pizarra.

Además se dispone de un proyector de cañón \emph{Epson Powerlite 77c}
de 2200 lúmenes que debe llevarse a clase cada vez. El profesor cuenta
con una netbook \emph{Lenovo IdeaPad s10e}, con procesador Intel Atom
de 1,60\,GHz, 1GB de memoria ram, 160GB de disco duro, sin unidad
óptica, dos puertos usb 2.0 y con pantalla de resolución $1024\times576$.
También se hizo una prueba con una computadora de la universidad,
una \emph{Toshiba Satellite A40} con procesador Mobile Intel Pentium
4 de 2,66\,GHz, 512MB de memoria ram (480MB + 32MB de video), 80GB
de disco duro, con unidad óptica combo CD-RW + lector de DVD, cuatro
puertos usb 1.1 y con pantalla de resolución $1024\times768$.

\begin{figure}
\begin{centering}
\includegraphics[width=1\columnwidth]{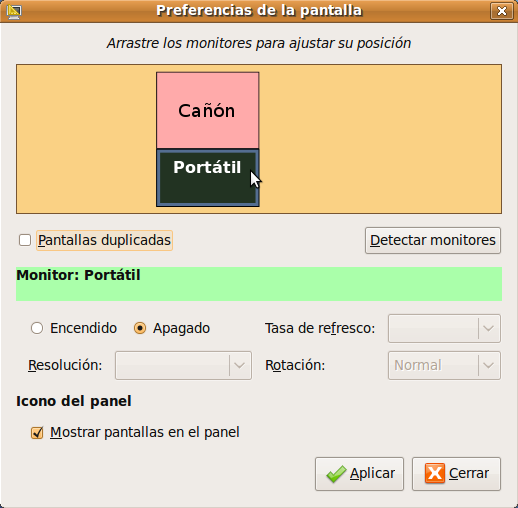}
\par\end{centering}
\caption{\label{fig:Arreglo-de-proyector-en-netbook}Arreglo de pantallas en
una netbook. La pantalla de la computadora está apagada y la pantalla
del proyector de cañón está encendida a $1024\times768$.}

\end{figure}

La computadora principal que se usa es una netbook, que tiene una
pantalla que no se puede configurar fácilmente a resolución de $1024\times768$,
el cañón no cuenta con la resolución de la computadora, de $1024\times576$,
y además, la aplicación de sincronización mencionado aquí no soporta
el uso de múltiples pantallas. Por ello hay que desactivar la pantalla
de la computadora y dejar sólo la imagen del proyector de cañón, tal
como muestra la figura \ref{fig:Arreglo-de-proyector-en-netbook}.
Con cualquier computadora que tenga resolución igual al cañón, $1024\times768$
o superior, simplemente hay que espejar las pantallas, es decir, que
proyecten la misma imagen con la misma resolución (que sería la opción
``Pantallas duplicadas'' en el cuadro de diálogo de la figura \ref{fig:Arreglo-de-proyector-en-netbook}).

Por otro lado, las dimensiones de la proyección y la distancia del
cañón a la superficie de proyección que se lograron y que se utilizaron
en clase fueron las siguientes (todas en $cm$):

\textbf{Escenario A:}\\
abajo: 97\\
arriba: 110\\
izquierda: 83\\
derecha: 86\\
proyector-superficie-abajo: 146\\
proyector-superficie-arriba: 182

\textbf{Escenario B:}\\
abajo: 137,\\
arriba 139,\\
izquierda: 103\\
derecha: 103\\
proyector-superficie-abajo: 209\\
proyector-superficie-arriba: 239

\pagebreak{}

\textbf{Escenario C:}\\
abajo: 153,\\
arriba 150,\\
izquierda: 110\\
derecha: 107\\
proyector-superficie-abajo: 230\\
proyector-superficie-arriba: 257

\begin{figure}
\begin{centering}
\includegraphics[width=1\columnwidth]{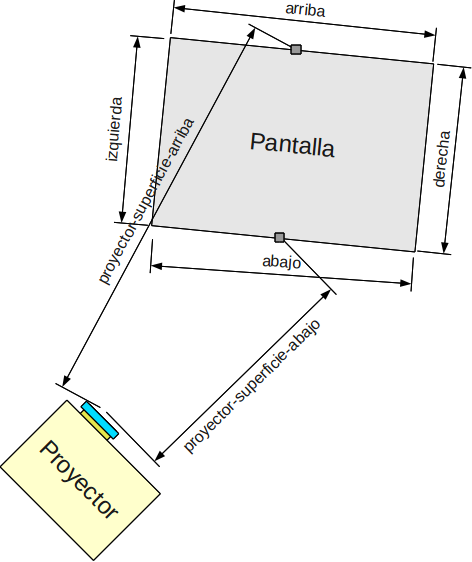}
\par\end{centering}
\caption{\label{fig:Dimensiones-logradas}Esquema de dimensiones relacionadas
con la proyección que se midieron durante las pruebas.}

\end{figure}

Donde \emph{abajo}, \emph{arriba}, \emph{izquierda} y \emph{derecha}
indican los lados de la superficie proyectada, y \emph{proyector-superficie-abajo}
y \emph{proyector-superficie-arriba}, indican la distancia entre la
lámpara del proyector y el punto medio del lado inferior y superior,
respectivamente, de la superficie proyectada (como en la figura \ref{fig:Dimensiones-logradas}).
En todos los casos, el wiimote se colocó justo encima del proyector
de cañón, como se aprecia en la figura \ref{fig:Wiimote-montado}.

Cabe mencionar que en algunos casos, las dimensiones de la superficie
de proyección no fueron mayores, debido a que el profesor ya no la
alcanzaba toda con el brazo extendido, y no debido a limitantes técnicas.

\section{Conclusiones y Resultados}

\subsection{Costos aproximados}

Los costos económicos aproximados de la solución presentada, suponiendo
que ya se cuenta con una computadora y con un proyector de cañón,
se presentan en el cuadro \vref{tab:Costos-aproximados}.

\begin{table}
\caption{\label{tab:Costos-aproximados}Costos aproximados}

\centering{}%
\begin{tabular}{|>{\raggedright}p{0.3\columnwidth}|>{\centering}p{0.25\columnwidth}|>{\centering}p{0.25\columnwidth}|}
\hline 
\textbf{Artículo} & \textbf{Costo mínimo} & \textbf{Costo máximo}\tabularnewline
\hline 
\hline 
\textcolor{blue}{Wiimote} & \textcolor{blue}{USD\,35 (usado)} & \textcolor{blue}{USD\,60 (nuevo)}\tabularnewline
\hline 
\textcolor{blue}{Pulsador infrarrojo casero (ver sección \ref{subsec:Construcci=0000F3n-pulsador})} & \textcolor{blue}{USD\,0,65} & \textcolor{blue}{USD\,4}\tabularnewline
\hline 
LED infrarrojo (como el del control remoto de la televisión) & USD\,0,30 & USD\,0,50\tabularnewline
\hline 
Cuvierta de cartón de nucleo de papel higiénico & USD\,0 & USD\,0\tabularnewline
\hline 
Revestimiento de cinta adhesiva & USD\,0 & USD\,0,50 (si se compra un rollo nuevo)\tabularnewline
\hline 
Portapilas para una pila AA & USD\,0,35 & USD\,1\tabularnewline
\hline 
Pulsador ``normalmente abierto'' & USD\,0 (si es desoldado de un ratón viejo) & USD\,1 (si es nuevo)\tabularnewline
\hline 
60\,cm de alambre para electrónica & USD\,0 (si es de un viejo cable UTP) & USD\,1\tabularnewline
\hline 
\textcolor{blue}{Espuma para maquetas de 5\,mm (ver el apéndice \vref{sec:Ap=0000E9ndice-plataforma})} & \textcolor{blue}{USD\,0 (si se consigue de residuos)} & \textcolor{blue}{USD\,3 (si se compra todo el pliego)}\tabularnewline
\hline 
\textbf{Total} & \textbf{USD\,35,65} & \textbf{USD\,67}\tabularnewline
\hline 
\end{tabular}
\end{table}

Otros costos posibles podrían incluir, alquilar un cautín (con la
respectiva pasta para soldar), unos 20\,cm de estaño y cinta aislante.
O en su defecto, contratar a alguien con más destrezas en electrónica
para que construya el dispositivo pulsador.

Si la computadora disponible no cuenta con un transmisor bluetooth
integrado, puede comprarse uno externo usb. Cuestan entre USD\,8
y USD\,15.

\subsection{Sobre la técnica}

Lo experimentado fue que mientras el wiimote se coloque sobre el proyector,
el sistema funciona, casi independientemente de la distancia hacia
la superficie de proyección. Las limitantes básicamente son dos: (a)
La estatura del profesor, ya que mientras más lejos estén entre sí
el cañón y la superficie de proyección, más grande es la imágen proyectada,
por lo que se requiere de más estatura para alcanzarla toda. (b) Mientras
más lejos esté el pulsador infrarrojo de la cámara del wiimote, más
débil es la señal que esta recibe, y deja de percibirse más allá de
5\,m (según las pruebas).

El montaje de todo el sistema se logra con eficiencia y práctica en
unos 15\,min (aunque se recomienda reservar 20\,min para las primeras
veces), habiendo hecho las pruebas pertinentes previamente. Esto incluye
conectar el proyector, encender una computadora portátil, armar el
accesorio de alineación (ver apéndice \ref{sec:Ap=0000E9ndice-plataforma})
y calibrar el wiimote sobre el proyector con las aplicaciones \texttt{wmgui}
y \texttt{gtkwhiteboard}. Es decir, todo el proceso descrito en la
sección \ref{subsec:Puesta-en-marcha} debería llevar no más de 20\,min
antes de la clase.

\subsection{Sobre los requerimientos}

En general los requerimientos, tanto de hardware como de software,
se mantienen sumamente bajos, lo que permite que esta implementación
se logre sin considerables inversiones de capital y que se logre con
hardware que la mayoría de usuarios de Windows considerarían obsoleto
en estos días. Esto incluye proyectores de cañón que tengan a lo sumo
resolución $1024\times768$\footnote{En realidad funciona con menos resolución, como $800\times600$ o
$640\times480$, pero son resoluciones poco cómodas para trabajar.} de proporción 4:3 y computadoras (laptops o desktops) que tengan
512MB de RAM y procesadores de la línea Pentium 4 o similares.

\subsection{Sobre la funcionalidad}

La funcionalidad de interactuar con el gestor de ventanas\footnote{Por decirlo de manera sencilla ``El Gestor de Ventanas'' es el componente
del sistema operativo encargado de dibujar las ventanas, sus bordes,
la barra de tareas, el escritorio, etc. Entonces, cuando se dice ``interactuar
con el gestor de ventanas'', se refiere a mover las ventanas, redimensionarlas,
maximizarlas, minimizarlas, seleccionarlas, etc.} a través del pulsador es eficiente, puesto que pueden realizarse
las tareas típicas: seleccionar menús, presionar botones, mover ventanas,
redimensionarlas, minimizarlas, restaurarlas, cerrarlas, etc. La precisión
es aceptable, y sólo depende de la habilidad (que viene con la práctica)
del profesor para utilizar el pulsador.

Sobre la velocidad y calidad de la escritura, eso depende de qué método
de escritura use el profesor:

\begin{figure}
\includegraphics[width=1\columnwidth]{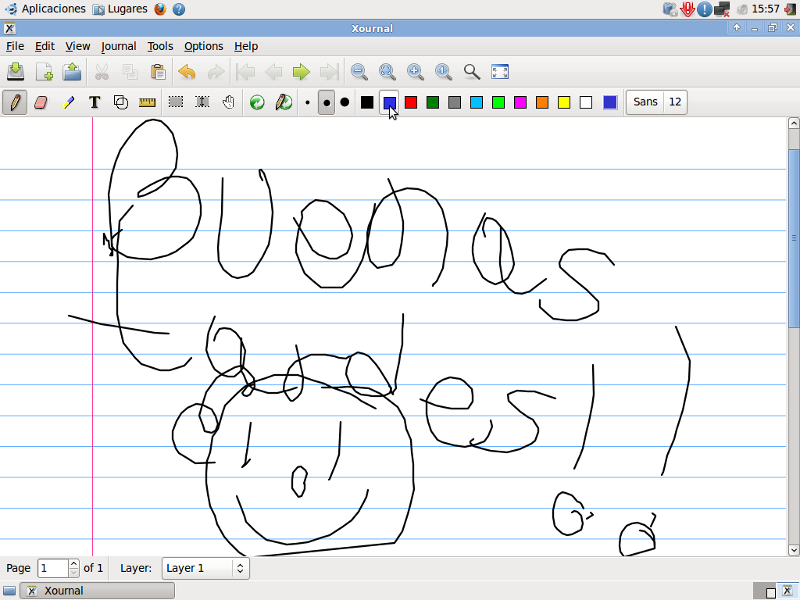}

\caption{\label{fig:Dibujo-de-trazos}Dibujo de trazos a mano alzada}

\end{figure}

\begin{figure}
\includegraphics[width=1\columnwidth]{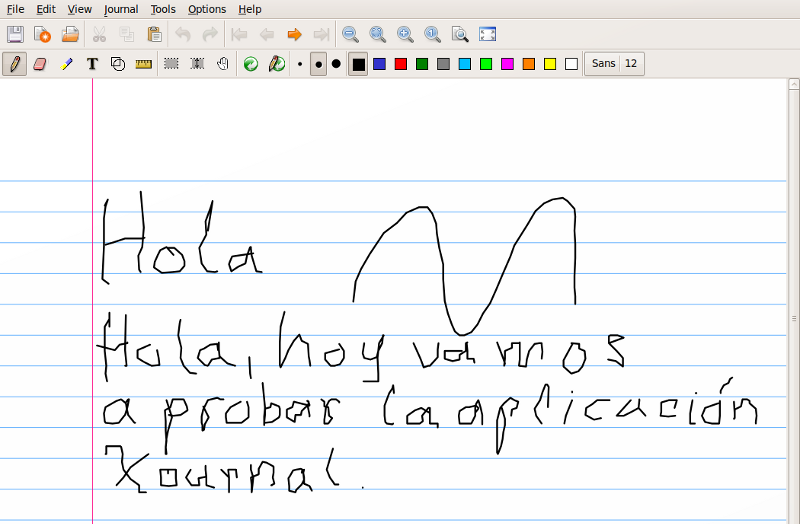}

\caption{\label{fig:Dibujo-de-texto}Escritura de texto a mano alzada}
\end{figure}

\begin{figure}
\includegraphics[width=1\columnwidth]{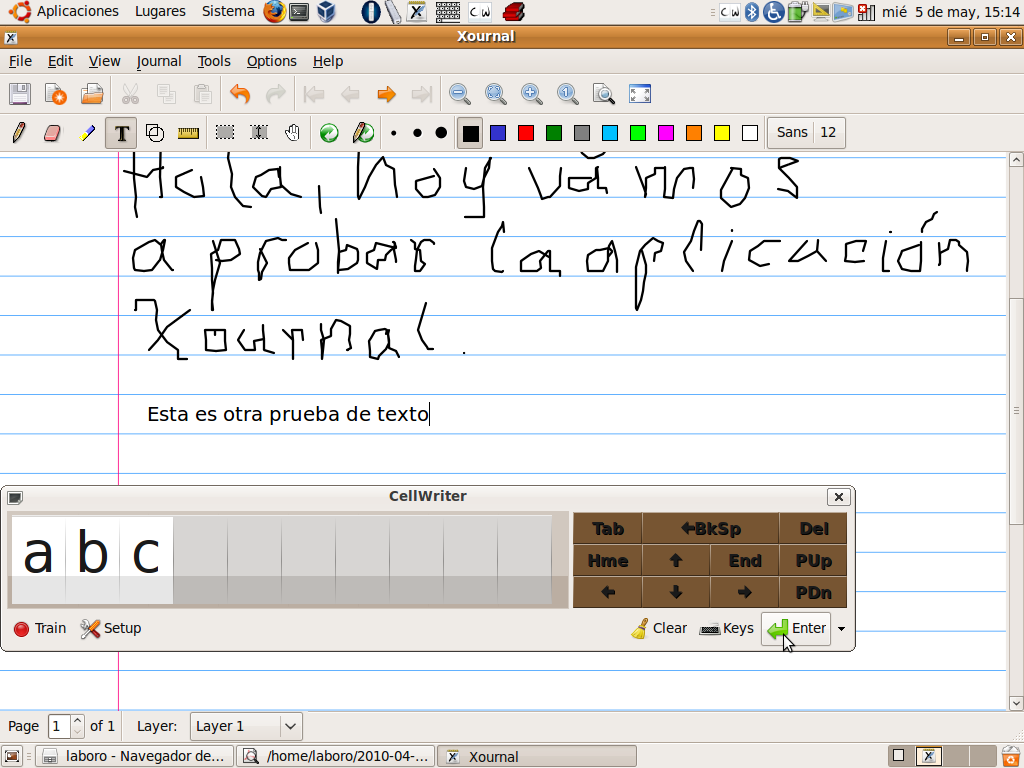}

\caption{\label{fig:Dibujo-de-texto-cellwriter}Dibujo de trazos a mano alzada}
\end{figure}

\begin{figure}
\begin{centering}
\includegraphics[width=0.9\columnwidth]{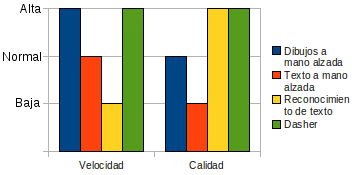}
\par\end{centering}
\caption{\label{fig:Comparaci=0000F3n-cualitativa-entre}Comparación cualitativa
entre diferentes métodos de escritura.}

\end{figure}

\begin{itemize}
\item Con dibujo de trazos a mano alzada, los resultados son muy buenos,
tanto en velocidad, como en calidad, como se ve en la figura \ref{fig:Dibujo-de-trazos}.
\item Con escritura de texto a mano alzada, la velocidad es la misma que
al escribir con yeso o con plumón en pizarras convencionales, pero
la calidad es bastante más baja, como se ve en la figura \ref{fig:Dibujo-de-texto}.
\item Si se utiliza algún programa de detección inteligente de escritura
(como \texttt{cellwriter}), la velocidad de escritura de texto se
reduce sensiblemente, pero la calidad obtenida es la de escribir en
el teclado de la computadora. Esto puede verse en la figura \ref{fig:Dibujo-de-texto-cellwriter},
la última línea (la que dice ``Esta es otra prueba de texto''),
se escribió con un programa de detección inteligente de texto que
se describe en la sección \ref{subsec:cellwriter}.
\item Si se utiliza el programa \texttt{dasher} (descrito en la sección
\ref{subsec:dasher}) diseñado para escritura en pantallas de PDA
y para escritura de personas con una o ninguna mano, la velocidad
de escritura de texto aumenta casi a la velocidad de escritura sobre
una pizarra convencional. Pero requiere de algo más de práctica que
los otros métodos.
\end{itemize}
La figura \ref{fig:Comparaci=0000F3n-cualitativa-entre} muestra una
comparación cualitativa entre los diferentes métodos de escritura
recién mencionados.

\subsection{En general}

Sobre la relación velocidad-calidad de escritura de texto, podemos
concluír de manera general con que: (a) La velocidad de escritura
es menor que con yeso o plumón. (b) La velocidad mejora con programas
como \texttt{dasher}. (c) La calidad mejora con programas como \texttt{cellwriter}.

El sistema es más práctico para aquellas circunstancias en las que
sea necesario dibujar mucho y escribir poco o hacer diagramas con
poco texto. En los casos en los que es necesario escribir mucho texto,
el resultado es una baja velocidad si se quiere texto con mucha calidad.

Si el texto ya se lleva a clase en forma de imagen o en forma de documento
PDF, y se utiliza este sistema para subrayar, marcar y/o enfatizar
el texto ya digitalizado, se pueden lograr mucho avance y mucha fluidez
en la clase.

Efectivamente se requiere de cierta experticia en programas de computadora
para adentrarse a utilizar este sistema sin recibir capacitación alguna.
Por lo que para que una persona pueda usarlo con fluidez y eficacia,
debería recibir una capacitación significativa sobre el tema.

Con algo de práctica --y dedicación-- el sistema puede ser un excelente
aliado para dinamizar las clases y para digitalizar el curso de las
mismas. Esto, por supuesto, depende del contenido del curso o materia,
y de la creatividad y la planificación del profesor encargado.

\section{Discución de los Resultados}

Podría ser más útil un pulsador con forma de varita mágica o regla\footnote{Podría construirse con un tubo vacío de micrófono estándar de computadora
o algo parecido.}, que tenga el led infrarrojo en la punta, a unos 15\,cm o 25\,cm
de la mano del profesor. Esto permitiría hacer una proyección más
grande que las mencionadas en la sección \ref{subsec:Prueba-real},
ya que el profesor tendrá mayor alcance con las manos.

Existe una implementación que también es libre: Wiimote Whiteboard
(\href{http://www.uweschmidt.org/wiimote-whiteboard}{http://www.uweschmidt.org/wiimote-whiteboard})
que está escrita en Java, especialmente para Mac. Aunque puede instalarse
en otros sistemas operativos, es más complicado de hacer, ya que requiere
de la instalación de la Máquina Virtual de Java y más software adicional
para poder correr la aplicación, como el .NET Framework 2.0 en Windows
que requieren más memoria RAM.

Por otro lado, si se pudiera construir un dispositivo equivalente
al wiimote, pero más barato, sería factible introducir esta implementación
como un producto de muy bajo costo entre las instituciones educativas.

\section{\label{sec:otros-programas}Apéndice sobre programas libres para
su uso en clases}

Todos los programas presentados aquí, son aplicaciones de código abierto
que pueden ser utilizados para diferentes propósitos en clases que
utilicen la implementación presentada por este artículo.

\subsection{xournal}

\emph{Xournal} es una aplicación concebida para tomar notas en pantallas
táctiles. Además, permite anotar sobre archivos PDF\cite{xournal}
y no precisa de clic derecho ni clic central para su uso, por lo que
es muy apropiada para nuestro propósito.

Las figuras \vref{fig:Dibujo-de-trazos} y \vref{fig:Dibujo-de-texto}
la muestran funcionando.

\subsection{\label{subsec:cellwriter}cellwriter}

\begin{figure}
\includegraphics[width=1\columnwidth]{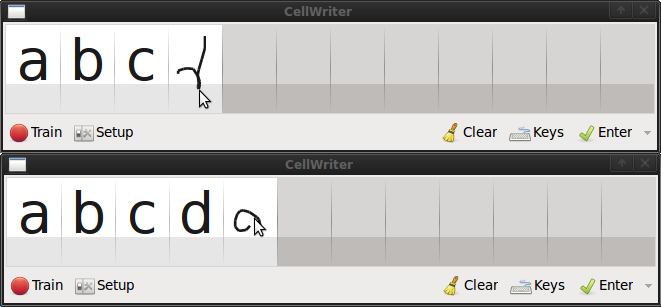}

\caption{\label{fig:Aplicaci=0000F3n-cellwriter}Aplicación \texttt{cellwriter}
que reconoce texto escrito a mano.}

\end{figure}

\emph{Cellwriter} es una aplicación de reconocimiento de escritura
a mano. Es fruto de una investigación en la Universidad de Minnesota.
Se basa en reconocimiento de patrones propios de la persona que escribe,
por lo que requiere que la persona entrene al programa, antes de poder
usarlo, para reconocer su manera propia de escribir. Puede enseñársele
a reconocer diferentes alfabetos de diferentes idiomas.\cite{cellwriter}

En la figura \ref{fig:Aplicaci=0000F3n-cellwriter} se muestra a la
aplicación antes y después de reconocer la letra ``d''. En nuestro
caso, la aplicación permite escribir a mano alzada con el pulsador
infrarrojo, y esta reconoce el texto como en la figura \vref{fig:Dibujo-de-texto-cellwriter}.

\subsection{\label{subsec:dasher}dasher}

\begin{figure}
\includegraphics[width=1\columnwidth]{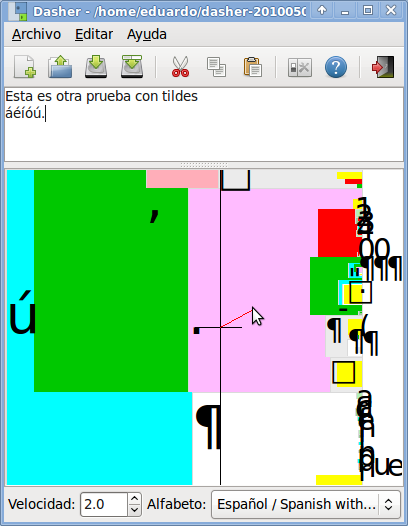}

\caption{\label{fig:Aplicaci=0000F3n-dasher}Aplicación \texttt{dasher} que
permite escribir a personas con diferentes discapacidades.}

\end{figure}

\emph{Dasher} es una aplicación de accesibilidad a computadoras para
personas que no pueden utilizar un teclado convencional para escribir.
Permite escribir texto únicamente moviendo el ratón o algún otro dispositivo
apuntador disponible. Además, aprende el patrón de uso de palabras
del usuario, por lo que a medida que más se use, más rápidamente se
obtienen las palabas que uno más frecuentemente usa.\cite{dasher}

Es una aplicación extraña a primera vista, y requiere algo de práctica
para entenderla, pero que en la práctica resulta muy útil para escribir
sin un teclado.

Puede verse la aplicación funcionando en la figura \ref{fig:Aplicaci=0000F3n-dasher}.
En nuestro caso es útil porque permite escribir simplemente moviendo
el pulsador infrarrojo sobre la superficie de proyección sin tener
que escribir a mano alzada, ni iniciar un teclado en pantalla. El
texto generado con esta aplicación puede fácilmente copiarse en \texttt{xournal}
por medio de un clic central en esta aplicación, el cual puede generarse
gracias a las ``zonas'' configuradas por \texttt{gtkwhiteboard}
(véase la figura \vref{fig:gtkwhiteboard}).

\subsection{OpenOffice.org}

\emph{OpenOffice.org} es la suite ofimática libre más desarrollada,
altamente compatible con Microsoft Office y un buen sustituto para
esta. Tiene aplicaciones para edición de documentos de texto, hojas
de cálculo, bases de datos locales, dibujo vectorial y creación de
presentaciones de diapositivas.\cite{openoffice}

\subsection{onboard}

Es la aplicación de teclado en pantalla del escritorio de Gnome, que
permite hacer uso del teclado, pero con el ratón.

\subsection{GIMP}

Es el programa de manipulación de imágenes GNU, que se puede utilizar
para hacer dibujos, trazos, retoque de fotos y otras manipulaciones
avanzadas de imágenes.\cite{gimp}

\section{Apéndice sobre costo de soluciones comerciales}

Existen muchas alternativas comerciales para una pantalla o pizarra
interactiva. Aquí presentamos algunas de ellas, para que se considere
el obstáculo económico que puede superarse con la solución presentada
en este artículo.

Las siguientes soluciones comerciales, se basan también en el wiimote:
\begin{itemize}
\item Sólo el programa de comunicación con el wiimote (una vez que se cancela
con tarjeta de crédito, se inicia la descarga del programa y su código
único de licencia privativa):

\begin{itemize}
\item \emph{Smoothboard} USD\,29,99\\
\href{http://www.smoothboard.net/}{http://www.smoothboard.net/} Este
sólo es el programa de comunicación con el wiimote, pero no es para
escribir en la pantalla.
\item \emph{iWiiBoard} USD\,29,99\\
\href{http://www.iwiiboard.com/}{http://www.iwiiboard.com/} Este
incluye la comunicación con el wiimote y también diversas herramientas
para escribir sobre el escritorio (hacer anotaciones).
\end{itemize}
\item Estos son unos Combos mixtos que incluyen diversos accesorios:

\begin{itemize}
\item Smoothboard Master Presenter's Pack USD\,239,95\footnote{A esto, hay que agregarle el costo de envío desde Estados Unidos}\\
Incluye: (2) Groove Pens (pulsadores infrarrojos ergonómicos y muy
bonitos), (2) plataformas con trípode de patas ajustables para wiimote,
(2) wiimotes, (1) transmisor bluetooth usb, (1) licencia del programa
Smoothboard, y dependiendo de ofertas temporales, una licencia del
programa \emph{ritePen 3.5}, que es para escribir en el escritorio
de windows.\\
\href{http://shop.irpensonline.com/}{http://shop.irpensonline.com/}
\item Smoothboard Dongle Edition USD\,79,99\\
Incluye: (1) CD que contiene una versión del programa Smoothboard
que corre desde el disco, sin instalarse sobre windows y (1) transmisor
bluetooth usb\\
\href{http://teachwithtech.com/Smoothboard-Dongle-Edition-p33.html}{http://teachwithtech.com/Smoothboard-Dongle-Edition-p33.html}
\end{itemize}
\item Pulsadores infrarrojos, de diversos modelos, de diversos precios,
9,95€, USD\,9,99, USD\,29,95, etc.\footnote{A esto, hay que agregar los costos de envío, y si sólo se compra un
artículo, el envío podría resultar más caro que el producto en sí.}:

\begin{itemize}
\item \href{http://www.wiiteachers.com/}{http://www.wiiteachers.com/}
\item \href{http://shop.infrawow.com/WIB-Scrib-001.htm}{http://shop.infrawow.com/WIB-Scrib-001.htm}
\item \href{http://www.penciil.com/}{http://www.penciil.com/}
\end{itemize}
\end{itemize}
La siguiente, es una solución completa: La \emph{Polyvision \={e}no
Next Generation Interactive Whiteboard}, una pizarra interactiva inalámbrica
(también se comunica por bluetooth) que también permite escritura
tradicional con plumón y además es magnética. Y según sus creadores,
es virtualmente indestructible. Trae software especial para usarla
y funciona con Windows y Mac.

Los costos identificados son\footnote{Y de nuevo, hay que considerar el costo de envío de algo tan grande.
Por otro lado, el autor encontró que se vende también en Panamá, pero
no logró obtener información de los costos.}: Alternativa 1, USD\,1595 en Estados Unidos, a través de \href{http://www.electronic-whiteboard.net/index/product/id/6763/}{http://www.electronic-whiteboard.net/index/product/id/6763/}.
Alternativa 2, desde USD\,1695 a USD\,1995 (dependiendo del tamaño)
en Estados Unidos, a través de: \href{http://techedu.com/}{http://techedu.com/}.

\section{\label{subsec:Construcci=0000F3n-pulsador}Apéndice sobre construcción
del pulsador infrarrojo}

\begin{figure}
\begin{centering}
\includegraphics[width=0.75\columnwidth]{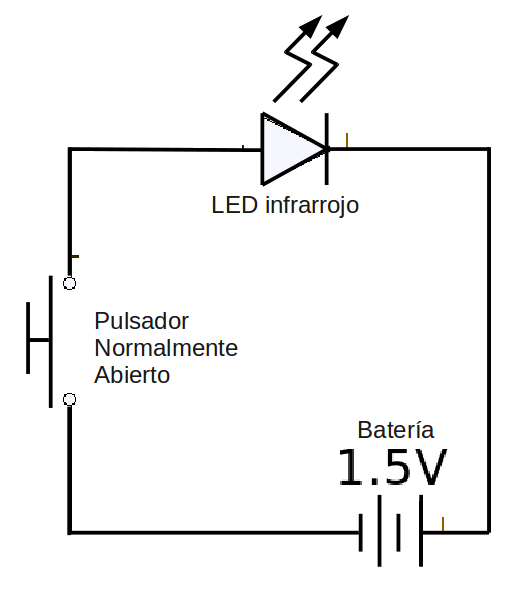}
\par\end{centering}
\caption{\label{fig:Diagrama-pulsador}Diagrama electrónico del pulsador infrarrojo}

\end{figure}

El pulsador infrarrojo es un dispositivo sencillo. Sus componentes
básicos son (véase la figura \vref{fig:Diagrama-pulsador}):
\begin{enumerate}
\item Un LED infrarrojo, como los que usan los controles remotos de televisores
y reproductores de DVD,
\item unos 60\,cm de alambre para electrónica,
\item un portabatería para una batería AA,
\item una batería AA,
\item un ``pulsador normalmente abierto'', es decir un interruptor que
cierre el circuito sólo mientras está presionado, y
\item alguna carcaza o compartimiento donde meter los componentes.
\end{enumerate}
Estos componentes pueden armarse de cualquier manera, siempre y cuando
su manipulación sea cómoda y funcional.

En el caso del autor, optó por meter los componentes en un tubito
de cartón, hecho con el cartón de un núcleo de rollo de papel higiénico
y sujeto con cinta adhesiva transparente. El pulsador normalmente
abierto fue extraído de un viejo ratón de bola ya descartado. No se
logró encontrar un portabaterías para una sola batería AA, por lo
que se usó uno para dos baterías AA y fue quebrado por la mitad con
una tenaza. Los alambres internos son fibras de un cable UTP (para
redes de computadora) descartado.

\begin{figure}
\begin{centering}
\includegraphics[width=1\columnwidth]{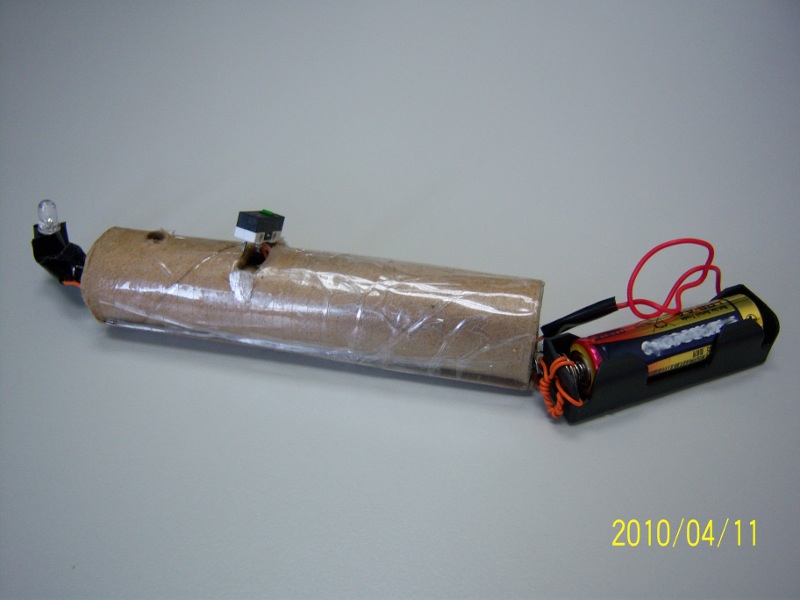}
\par\end{centering}
\caption{\label{fig:Pulsador-desarmado}Pulsador con las partes internas expuestas}

\end{figure}

El dispositivo sin terminar puede verse en la figura \ref{fig:Pulsador-desarmado}.

Los componentes deben conectarse y soldarse según el diagrama de la
figura \ref{fig:Diagrama-pulsador}, introducirse en el compartimiento
elegido y permitir sustituír la batería, cuando esta se agote. Algo
muy importante a considerar es que la dirección de emisión de la luz
del LED debe ser, en la medida de lo posible, hacia la cámara infrarroja
del wiimote para que pueda verla siempre sin problemas. Por eso, el
dispositivo mostrado arriba tiene el LED en dirección perpendicular
al resto del dispositivo.

\begin{figure}
\begin{centering}
\includegraphics[width=0.8\columnwidth]{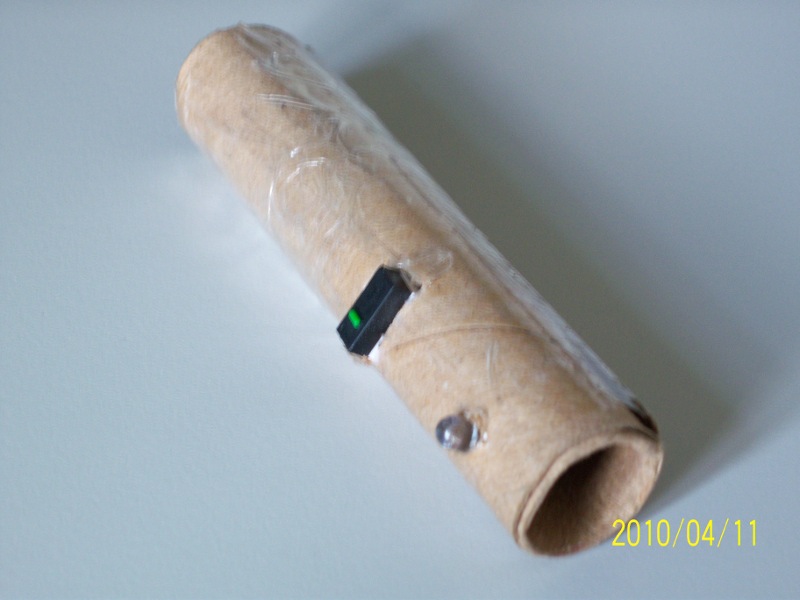}
\par\end{centering}
\caption{\label{fig:Pulsador-terminado}Pulsador terminado}
\end{figure}

La apariencia final del dispositivo se muestra en la figura \ref{fig:Pulsador-terminado}.
Por supuesto, el dispositivo puede armarse de cualquier manera, y
construírse de cualquier tipo de materiales.

\section{\label{sec:Ap=0000E9ndice-plataforma}Apéndice sobre construcción
de accesorio de alineación para facilitar el montaje}

Para que esta implementación funcione, el wiimote debe permanecer
estático durante su uso en las clases. Puede utilizarse cualquier
artilugio para lograrlo. Esto incluye pedestales de micrófono, pedestales
de cámara, papeles, libros, etc. Incluso se puede implementar una
solución muy ecológica, con una bolsa impermeable llena de arena o
de harina, tal que mantenga su forma con el poco peso del wiimote
encima.

Una característica muy deseable es que funcione como aislante térmico
entre el calor generado por el proyector y el wiimote.

En este apartado se presenta una solución económica, fácil de transportar
y elegante. Se requiere de unos 30cm$\times$30cm de espuma para hacer
maquetas, o cualquier otro material relativamente fácil de cortar,
liviano y más resistente que el cartón.

En este caso, se utilizó una espuma forrada para hacer maquetas (\emph{Foam
Board}) de 5mm de espezor.

\begin{figure}
\includegraphics[width=1\columnwidth]{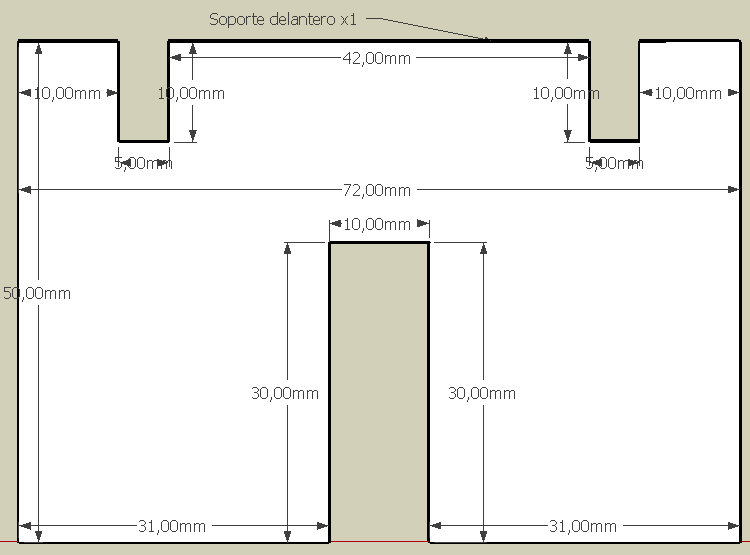}

\caption{\label{fig:Planos-plataforma-1}Pieza - Soporte delantero - 1x}
\end{figure}

\begin{figure}
\includegraphics[angle=90,width=0.85\columnwidth]{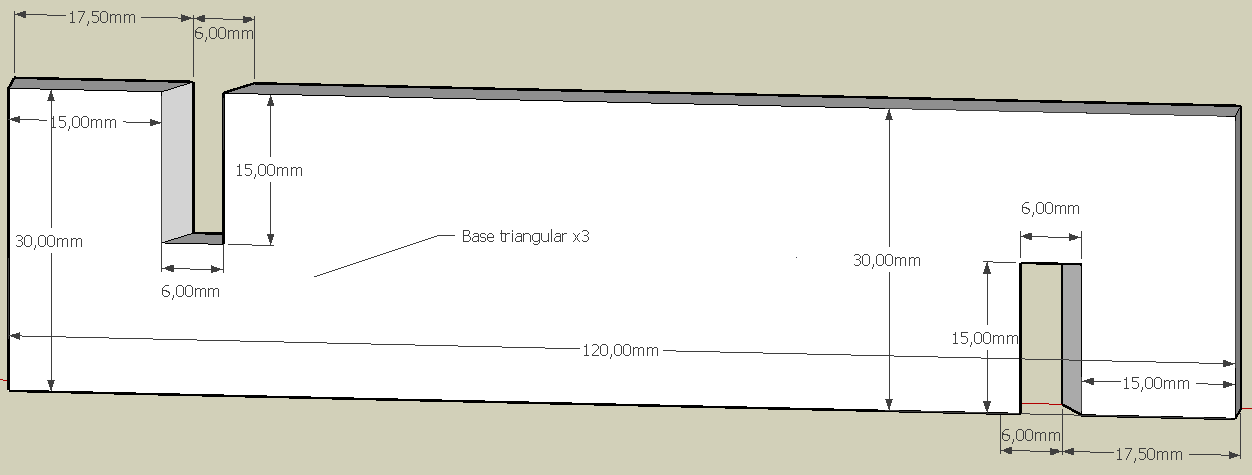}

\caption{\label{fig:Planos-plataforma-2}Pieza - Base triangular - 3x\protect \\
Las ranuras de anclaje deberían idealmente tener una inclinación de
60º.}
\end{figure}

\begin{figure}
\includegraphics[angle=90,width=0.75\columnwidth]{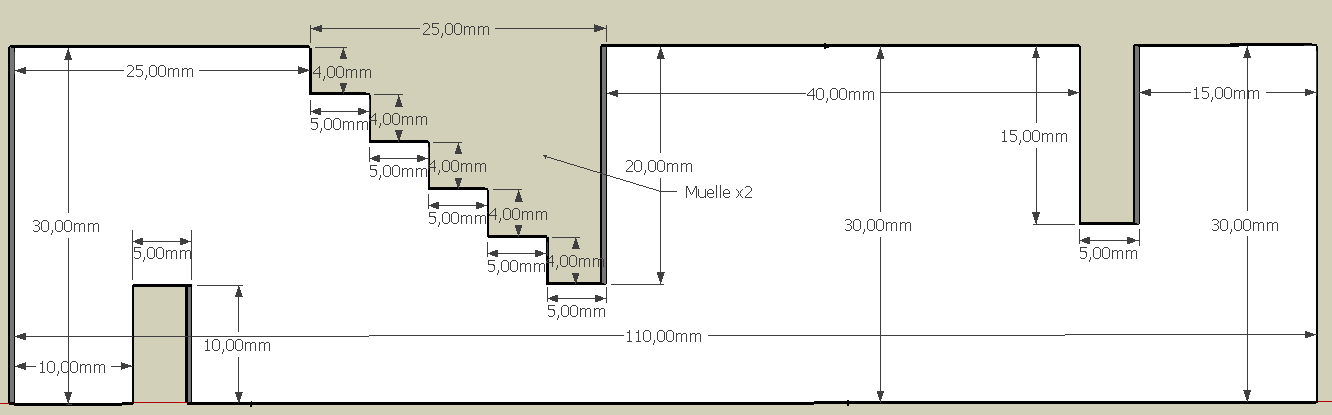}

\caption{\label{fig:Planos-plataforma-3}Pieza - Muelle - 2x}
\end{figure}

\begin{figure}
\includegraphics[angle=90,width=0.8\columnwidth]{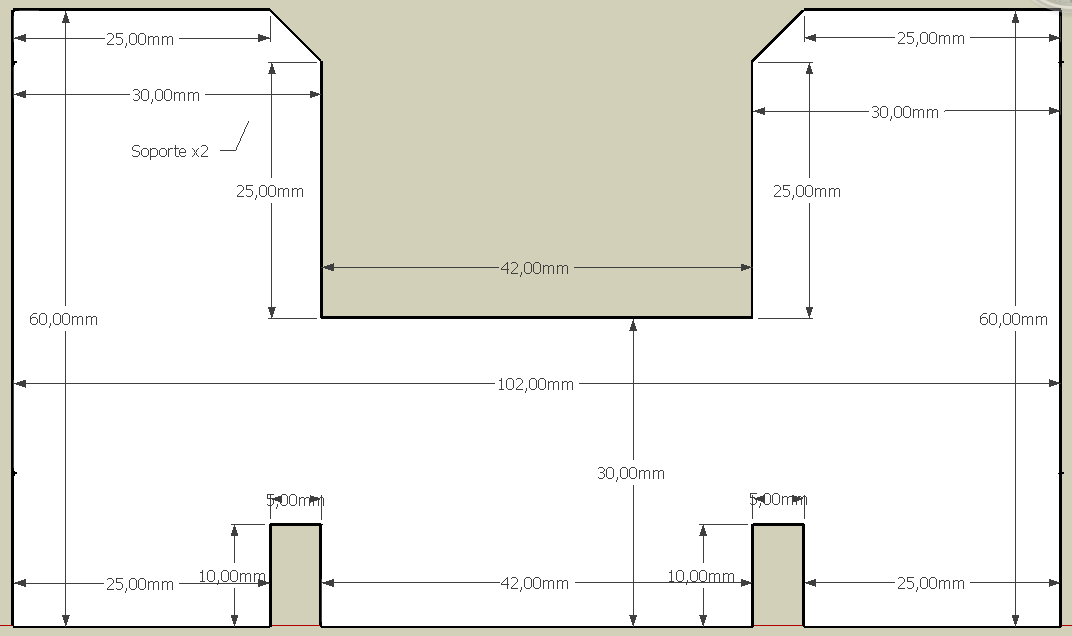}

\caption{\label{fig:Planos-plataforma-4}Pieza - Soporte - 2x}
\end{figure}

\begin{figure}
\begin{centering}
\includegraphics[width=0.8\columnwidth]{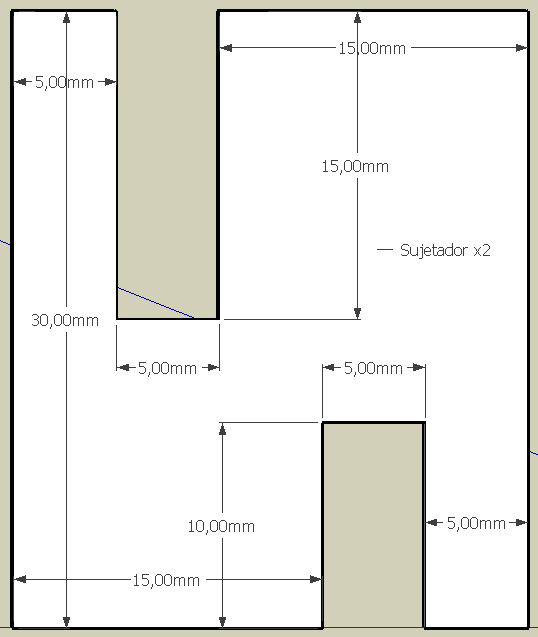}
\par\end{centering}
\caption{\label{fig:Planos-plataforma-5}Pieza - Sujetador - 2x}
\end{figure}

\begin{figure}
\includegraphics[width=0.49\columnwidth]{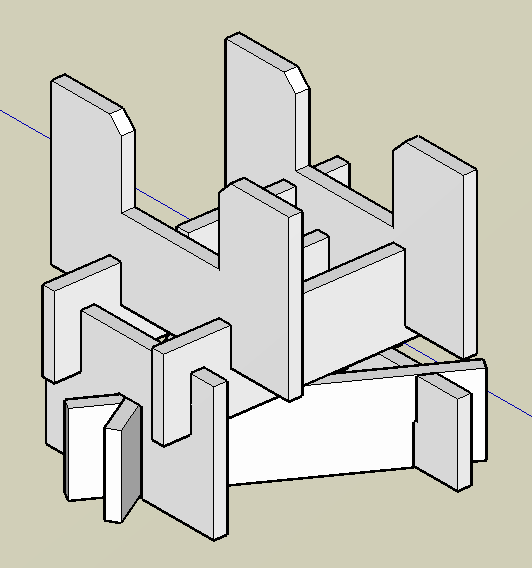}\hfill{}\includegraphics[width=0.49\columnwidth]{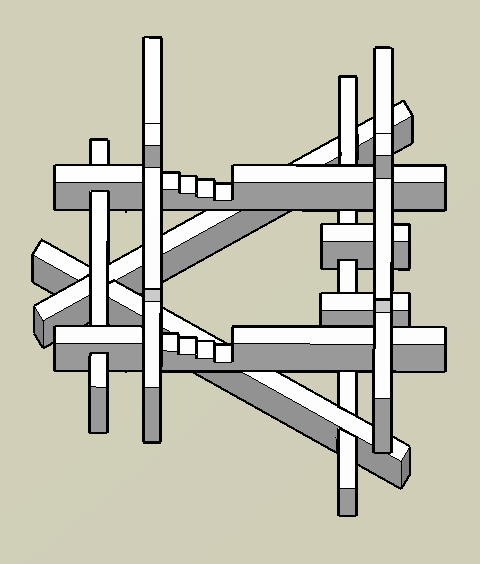}

\caption{\label{fig:Plataforma-armada}Plataforma armada, en vista frontal
y vista aérea.}

\end{figure}

\begin{figure}
\begin{centering}
\includegraphics[width=1\columnwidth]{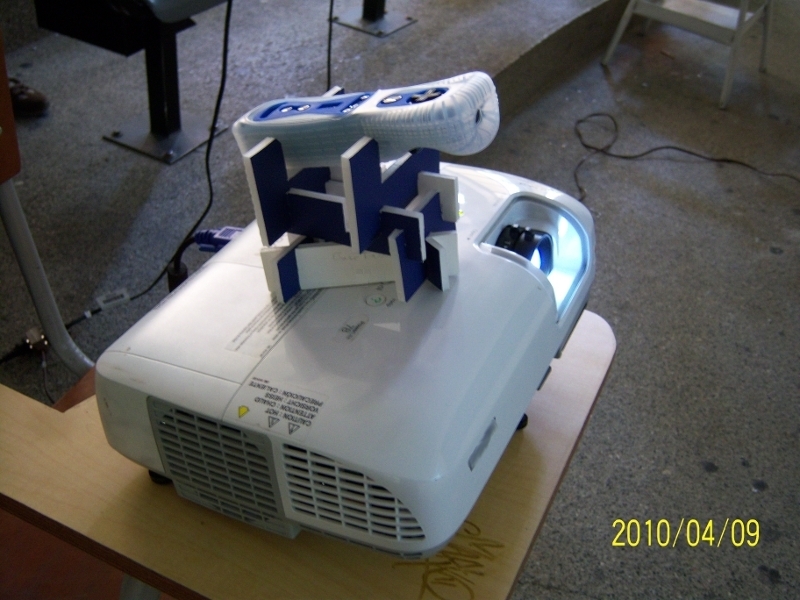}
\par\end{centering}
\caption{\label{fig:Wiimote-montado}Wiimote montado sobre un proyector de
cañón, con el accesorio de alineación}

\end{figure}

Las figuras \ref{fig:Planos-plataforma-1}, \ref{fig:Planos-plataforma-2},
\ref{fig:Planos-plataforma-3}, \ref{fig:Planos-plataforma-4} y \ref{fig:Planos-plataforma-5}
muestran los planos de la plataforma propuesta. Las figuras \ref{fig:Plataforma-armada}
y \ref{fig:Wiimote-montado} muestran la plataforma construída digitalmente
y construída en la realidad, respectivamente.

Este diseño cumple con los requerimientos de ser lo suficientemente
fuerte para sostener el wiimote, de ser fácil de transportar y rápido
de montar. Además, mantiene una distancia considerable entre el control
y el proyector, disminuyendo la transferencia de calor del segundo
al primero.

\section{Reconocimientos}

Gracias a Johnny Chung Lee, constructor del primer programa que aprovecha
el hardware del wiimote para usarlo con computadoras.\cite{chung-lee}

Gracias a Williams Roque, inspirador de esta tecnología en el autor.

Y por último, pero no menos importante, se le agradece grandemente
al Lic. Guillermo Cortés, Coordinador de la Carrera de Licenciatura
en Ciencias de la Computación, por su dedicación al revisar este artículo
e identificar detalles importantes que el autor había pasado por alto.

\end{document}